\documentclass[oldversion]{aa}

\usepackage{epsfig}
\usepackage{graphics}
\usepackage{float}
\usepackage{amsmath}
\usepackage{multirow}
\usepackage{longtable}
\usepackage{rotate}
\usepackage{array}
\usepackage{subfigure}
\def\kms{\ifmmode{\rm km\,s^{-1}}\else\hbox{$\rm km\,s^{-1}$}\fi}

\setlongtables

\begin{document}

\title{Mass estimates for visual binaries with incomplete orbits}

\author{L.B.Lucy}
\offprints{L.B.Lucy}
\institute{Astrophysics Group, Blackett Laboratory, Imperial College 
London, Prince Consort Road, London SW7 2AZ, UK}
\date{Received ; Accepted }

\abstract{The problem of estimating the total mass of a visual binary when its
orbit is incomplete is treated with Bayesian methods.
The posterior mean of a mass estimator is 
approximated by a triple integral over orbital period, orbital eccentricity 
and time of 
periastron.
This reduction to 3-D from the 7-D space defined by the 
conventional Campbell parameters is achieved by adopting
the Thiele-Innes elements and exploiting the linearity with respect to the
four Thiele-Innes constants. 
The formalism is tested on synthetic observational data
covering a variable fraction of a model binary's orbit.
The posterior mean of the mass estimator 
is numerically found to be unbiased when the data cover $\ga 40\%$ of the orbit.
}

\authorrunning{Lucy}
\titlerunning{Visual binaries}
\maketitle

\section{Introduction}

Visual binaries are a fundamental source of data on
stellar masses. An individual system provides definitive data if
it has a precise parallax $\varpi$ {\em and} 
a high quality 
orbit yielding accurate values for the orbital period $P$ and the semi-major
axis $a$. The total mass of the binary in solar units is then given by
Kepler's third law
\begin{equation} 
 {\cal M}_{1} + {\cal M}_{2}  = \frac{1}{\varpi^{3}} \: \frac{a^{3}}{P^{2}} 
\end{equation}
where $P$ is in years and $a$ and $\varpi$ are in arcseconds.     
However, for many long-period binaries,
the accumulated measurements do not cover even one
full orbit.  The question therefore arises: can a high quality 
orbit be derived from such incomplete data?   

In his classic monograph, Aitken (1918) gave the following
answer: 'In general, it is not worthwhile to compute the orbit of a
double star until the observed arc not only exceeds 180\degr, but also defines 
both ends of the apparent ellipse.' Since the first observation 
will seldom coincide with either end point, Aitken's criterion
typically requires $f_{orb} \ga 0.75$, where
$f_{orb}$ is the fraction of the orbit covered. 
Aitken also warns that when $f_{orb} < 0.5$ 
it will generally be possible
to draw several very different ellipses each of which will satisfy the data 
about equally well.

Although definitive orbits are surely a prerequisite for definitive masses,
Eggen (1967) noted that $a^{3}/P^{2}$ is often reasonably well
determined even when $a$ and $P$ are not, a conclusion he reached by
compiling data on systems with multiple computed orbits. He reports
(Eggen 1962) that 'In some cases there have been drastic revaluations of
the period of a given pair but there is then a compensating change in $a$
and the resulting values of $a^{3}/P^{2}$ suffer little alteration.'   
These same compensating changes in $a$ and $P$ also arise in the Monte Carlo
search technique of Schaefer et al. (2006) - see their Fig.13 for DF Tau.

Eggen's examination of the historical record provides strong empirical
evidence that useful masses can be obtained from incomplete orbits.
But restricting the discussion to {\em observed} systems does not
allow an assessment of the statistical properties of such mass estimates.
Accordingly, in this paper, synthetic observations with $f_{orb} < 1$
are created for a binary with specified elements in order
to test the accuracy of the derived total mass.

The problem of inferring masses from incomplete orbits will be particularly
acute for data acquired by the {\em Gaia} satellite. Nurmi (2005)
and Pourbaix (2011) estimate that $\ga 10^{7}$ visual binaries
will be discovered. Many of these will have periods 
longer than the five years planned for the mission, and their orbits
could perhaps be analysed by the technique developed in this paper.

\section{Bayesian estimation}

As noted already by Aitken (1918), when an orbit is incomplete,
several theoretical orbits will provide satisfactory fits.
Two possible responses to this circumstance are: 1) scan parameter space 
to find the global minimum (e.g., Hartkopf et al. 1989)
and assume that this orbit is the closest achievable approximation to the truth; 
or 2) scan parameter space to make a complete census of acceptable orbits and 
then compute an appropriate average of the individual mass determinations
(e.g., Schaefer et al. 2006). 
However, a Bayesian
approach is preferred here. Specifically, by scanning parameter space,
the posterior means of the orbital elements or any function thereof can be
computed without locating  minima.

\subsection{Orbital elements} 

The standard orbital elements describing a  
secondary's motion relative to its primary are the Campbell elements 
$\theta = (P,T,e,a,{\rm i},\omega,\Omega)$. In addition to 
$P$ and $a$ previously defined, $T$ is a time of periastron passage, $e$
is the eccentricity, ${\rm i}$ is the inclination, $\omega$ is the longitude of 
periastron, and $\Omega$ is the position angle of the ascending node.
However, for scanning parameter space, it is convenient to define an
alternative to $T$.
Noting that $T$ can be replaced by $T \pm nP$, where $n$ is any integer,
we choose $T \in (0,P)$ and then define the alternative element   
$\tau = T/P \in (0,1)$.

The above Campbell elements are standard. But an alternative set,
the Thiele-Innes elements (Appendix A), simplify Bayesian integrations.
In these elements, the vector $(a,{\rm i},\omega,\Omega)$ is replaced by 
$\psi = (A,B,F,G)$  , so that the
7-element vector of elements is now $(\phi, \psi)$, where
$\phi = (P,e,\tau)$.

Numerous authors have preferred the Thiele-Innes elements for conventional 
least-squares determinations of orbital parameters for both visual binaries
and exoplanets (e.g., Hartkopf et al. 1989, Pourbaix et al. 2002, 
Schaefer et al. 2006, Casertano et al. 2008, Wright \& Howard 2009). 
This paper argues that these elements are also advantageous for Bayesian
estimation of orbital parameters.

\subsection{Posterior mean} 

Let $Q(\theta)$ be a quantity whose value would normally
be computed after deriving a definitive orbit. But now in anticipation
of there being no single definitive orbit, we instead compute $Q$'s posterior
mean, given by
\begin{equation}
 <\!Q\!> \: = \int{Q \: {\cal L} \: p \: d \phi d \psi} \:
                      / \int{{\cal L} \: p \: d \phi d\psi}
\end{equation}
Here ${\cal L}(\phi, \psi| D)$ is the likelihood of the elements 
$(\phi, \psi)$ given the data $D$, and $p(\phi, \psi)$ is a probability
density function (pdf) quantifying our prior beliefs or knowledge
about the orbital elements.
In this problem, $D$ comprises the secondary's measured relative 
Cartesian sky coordinates $(\tilde{x}_{n},\tilde{y}_{n})$ at times 
$t_{n}$.

As remarked in Sect.2.1, the Thiele-Innes elements simplify the
integrations in Eq.(2). 
For fixed
$\phi = (P,e,\tau)$, the family of Keplerian orbits is {\em linear} in
$A,B,F,G$. Accordingly, their 
least-squares values $\hat{A},\hat{B},\hat{F},\hat{G}$ can be computed without
iteration - see Sect.A.2. (Note that with normally-distributed errors   
the least-squares solution is also the point of maximum likelihood (ML).)

The second simplification concerns topology.
When $f_{orb} \la 0.5$, several orbits fit the data (Sect.1). This implies
severe topological complexity for ${\cal L}$ in the 7-D 
$(\phi,\psi)$-space. 
However, at each point in $\phi$-space, the 4-D function 
${\cal L}(\psi) / {\cal L}(\hat{\psi})$ 
is a quadrivariate
normal distribution (Sect.A.3), which therefore has a
{\em single} peak at 
$\hat{\psi} = (\hat{A},\hat{B},\hat{F},\hat{G})$, the point of ML. 

This topological simplification can be used in estimating 
the 7-D integrals in Eq.(2). To achieve this, we approximate the
quadrivariate normal distribution by the delta function 
$\delta(\psi - \hat{\psi})$, so that
\begin{equation}
 {\cal L}(\phi, \psi | D) =  {\cal L}^{\dag}  \: \delta(\psi - \hat{\psi}) 
\end{equation}
where   
\begin{equation}
 {\cal L}^{\dag} =  {\cal L} (\phi, \hat{\psi} | D) 
\end{equation}
Thus, for each point in $\phi$-space,  ${\cal L}^{\dag}$ is the value of 
${\cal L}$ at $\hat{\psi} = \hat{\psi}(\phi)$, the ML
point in $\psi$-space. 

If we now replace  ${\cal L}$ in Eq.(2) by the approximation given in Eq.(3)
and integrate over $\psi$-space, we obtain
\begin{equation}
 <\!Q\!> \: = \int Q^{\dag} \: {\cal L}^{\dag} \: p^{\dag} \: d \phi \:  
                / \int{\cal L}^{\dag} \: p^{\dag} \: d \phi 
\end{equation}
where the superscript $\dag$ indicates evaluation at $(\phi,\hat{\psi})$
as in Eq.(4).
 
On the assumption of normally-distributed measurement errors and with 
constants of proportionality omitted,  
\begin{equation}
 {\cal L}^{\dag} = \exp \: ( - \frac{1}{2} \chi^{2} )
\end{equation}
where  $\chi^{2}(\phi,\hat{\psi})$ is the conventional goodness-fit 
criterion - see Eq.(A.5) - at $\phi = (P,e,\tau)$.

In the 
statistics literature ${\cal L}^{\dag}$ is the {\em profile
likelihood} (e.g., Severini 2000). This function is 
used in high energy physics for evaluating the significance of particle 
detections (e.g., Ranucci 2012).

\subsection{Priors}

In order to calculate the approximate posterior mean $<\!Q\!>$ from 
Eq.(5), $p^{\dag} = p(\phi,\hat{\psi})$ must be specified.
Guided by common practice in 
Bayesian exoplanet detection (e.g., Ford \& Gregory 2007), we assume 
the elements to have {\em independent} priors, so that $p^{\dag}$ is the 
product of seven independent functions. The priors on the $\psi$ elements
are taken to be uniform
and so cancel between numerator and denominator in Eq.(5).  

For the $\phi$ elements $e$ and $\tau$, the chosen priors are uniform 
in $(0,1)$. For $P$, 
we adopt a Jeffreys prior - i.e., $\log P$ uniform in $(\log P_{L},\log P_{U})$.

\subsection{Numerical integration}

The integrals in Eq.(5) are 3-D as against 7-D in Eq.(2).
This reduction allows $<\!Q\!>$ to be evaluated simply by computing the 
integrands at every point in a 3-D grid 
and then summing. Without this reduction,
a Markov Chain Monte Carlo (MCMC) calculation would perhaps be required. 
Direct summation would, however,
still be feasible if regions of negligible likelihood were efficiently excluded
(Mikkelsen et al. 2012).

The integration domain in $(\log P,e,\tau)$-space is taken to be
$(\log P_{L},\log P_{U})$ for $\log P$, and $(0,1)$ for both $e$ and $\tau$.  
This domain is partitioned into a 3-D grid with constant steps for each
variable.
The quantity $Q$ is then evaluated at the mid-points of the
cells labelled $(i,j,k)$ with $i \in (1,I)$, $j \in (1,J)$, $k \in (1,K)$.
The resulting formula for the posterior mean is 
\begin{equation}
 <\!Q\!> = \Sigma_{ijk} \: Q_{ijk}^{\dag} \: {\cal L}_{ijk}^{\dag}
                          / \: \Sigma_{ijk} \: {\cal L}_{ijk}^{\dag}
\end{equation} 
The priors adopted in Sect.2.3 are implicitly incorporated 
since the cells are weighted equally.

\section{Calculation of $<\!Q\!>$}

In this section, the calculation of $<\!Q\!>$ is described step-by-step.

\subsection{Model binary}

The model binary has the following Campbell elements:
\begin{eqnarray}
  P_{*}=100y  \;\;\; \tau_{*}=0.4 \;\;\; e_{*}=0.5  \;\;\; a_{*}=1\arcsec  
                                                   \nonumber    \\
  {\rm i}_{*}=60\degr   \;\;\;    \omega_{*} = 250\degr    \;\;\;  
                                              \Omega_{*} = 120\degr
\end{eqnarray}
From Eqs.(A.1), the Thiele-Innes elements corresponding to the above values
of $a_{*},{\rm i}_{*},\omega_{*},\Omega_{*}$ are
\begin{eqnarray}
     A_{*} = +0\farcs578  \;\;\;\;\;   B_{*} = -0\farcs061   \nonumber    \\
     F_{*} = -0\farcs322  \;\;\;\;\;   G_{*} = +0\farcs899
\end{eqnarray}
\subsection{Synthetic data}

Given these elements, the binary is 'observed' at times
\begin{equation}
 t_{n} = f_{orb} P_{*}   \times (n-1)/(N-1)  
\end{equation}
for $n = 1,2,  \dots, N$, so that the fraction $f_{orb}$ of the orbit is
uniformly sampled. At each $t_{n}$, the secondary's exact coordinates 
$(x_{n},y_{n})$ are computed from Eqs.(A.2)-(A.4). The
measured positions are then   
\begin{equation}
 \tilde{x}_{n} =  x_{n} + \sigma \: z_G   \;\;\;\;
                          \tilde{y}_{n} =  y_{n} + \sigma \: z_G 
\end{equation}
where each $z_G$ is a random gaussian variate drawn from ${\cal N}(0,1)$
and $\sigma$ is the standard error of unit weight.
The vector $(f_{orb},N,\sigma)$ defines an observing campaign.

Uniform sampling is here assumed for simplicity: it is not required by the
Bayesian technique. Random sampling, the obvious alternative, is an 
inferior model for extant ground-based data on long-period binaries where 
typically an observation was made every observing season.

\subsection{Calculation procedure}

Given positions $(\tilde{x}_{n},\tilde{y}_{n})$,
a posterior mean $<\!Q\!>$ is computed as follows:

1) The grid parameters $I,J,K$ and $P_{L},P_{U}$ are specified.  

2) At each grid point, the vector $\phi_{ijk} = (P_{i},e_{j},\tau_{k})$ 
follows from the stepping procedure described in Sect.2.4.

3) Given $\phi_{ijk}$, the least-squares (ML)
values of the Thiele-Innes elements 
$\hat{\psi} = (\hat{A},\hat{B},\hat{F},\hat{G})$ are computed from Eq.(A.7).

4) Given $\hat{\psi}$, the predicted positions $(x_{n},y_{n})$
are computed from Eqs.(A.2)-(A.4) 
and the resulting $\chi^{2}_{ijk}$ derived from Eq.(A.5).

5) Given $\chi^{2}_{ijk}$, the profile likelihood 
${\cal L}^{\dag}_{ijk}$ is given by Eq.(6).  

6) Given $\hat{\psi}$, the corresponding Campbell elements 
are computed from Eqs.(A.11)-(A.15) and used to evaluate 
$Q_{ijk} = Q(\theta_{ijk})$.

7) Given $Q_{ijk}$ and ${\cal L}^{\dag}_{ijk}$ throughout the grid,
$<\!Q\!>$ is obtained from Eq.(7).

Note that steps 1) - 4) also arise in the parameter search techniques of 
Hartkopf et al. (1989) and Schaefer et al. (2006).

\section{Feasible orbits}

In this section, we exploit the scanning of parameter space required by
the calculation of $<\!Q\!>$ to relate the findings of Aitken and
Eggen (Sect.1) to the behaviour of ${\cal L}^{\dag}(\phi|D)$.

\subsection{A particular case}

With orbital parameters from Eq.(8) and campaign parameters 
$f_{orb} = 0.4, N = 15, \sigma = 0\farcs05$, synthetic data are calculated from
Eq.(11). Then, with grid parameters 
$P_{L} = 0.7 f_{orb} P_{*}$, $P_{U} = 300 P_{*}$, $I = 800, J = K = 200$,   
the values $\chi^{2}_{ijk}$ at all grid points
are determined as described in Sect.3.3.

If at grid point $(i,j,k)$,
\begin{equation}
   P(\chi^{2} > \chi^{2}_{ijk}) > 0.05 
\end{equation}
the elements $\theta_{ijk}$ are
deemed to represent a {\em feasible} orbit, and the ensemble of such
orbits define the feasible domain(s) ${\cal D}$ in $(\log P, e, \tau)$-space.

In Fig.1, ${\cal D}$ 
is projected onto the $(\log P, e)$ plane. Thus a filled circle appears
at the point $(\log P_{i}, e_{j})$ if Eq.(12) is fulfilled for at least one
$\tau_{k}$. From this plot, we see that for this poorly-observed incomplete
orbit, there is an extended domain of feasible orbits with $P$'s
ranging from $40$ to $5000y$ and $e$'s from $0.19$ to $> 0.99$. 

Because  ${\cal D}$ extends to the grid's upper boundary at $e = 1$, a
more general treatment would, in this particular case, show that some parabolic
and even hyperbolic orbits are feasible - i.e., fit the data. By
restricting the analysis to elliptical orbits, unbound orbits are
excluded. But this is appropriate for the practical problem 
of concern: an incomplete observed orbit is highly unlikely to be a close
encounter of an unbound pair.

\begin{figure}
\vspace{8.2cm}
\includegraphics{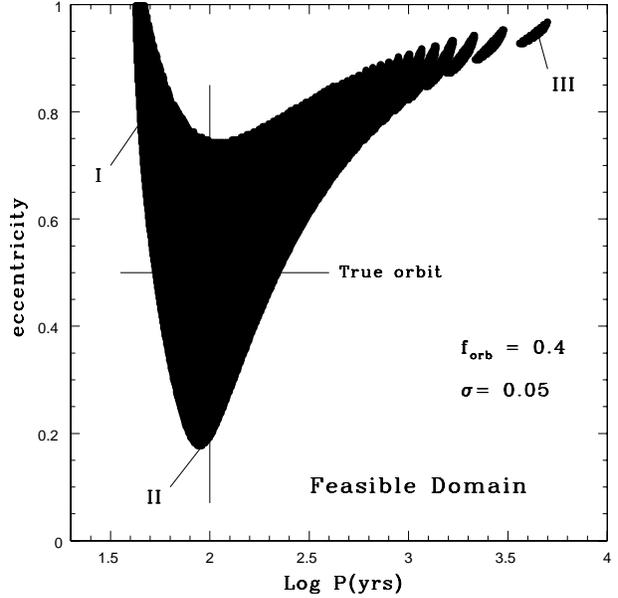}
\caption{Feasible domain ${\cal D}$ in $(\log P, e)$-space for the test 
binary defined
by Eq.(8) when observed in a 
campaign with parameters $f_{orb} = 0.4, N = 15, 
\sigma = 0\farcs05$. The true values $\log P_{*}, e_{*}$ are indicated
as are the coordinates of the orbits plotted in Fig.2.}
\end{figure}

In Fig.2, the true orbit for this particular case is plotted together with
the measurements $(\tilde{x}_{n},\tilde{y}_{n})$. 
Also shown are three feasible orbits indicative of the range seen in 
Fig.1. The Campbell elements of these are: 
\begin{eqnarray}
  P=47.0y  \;\;\; T = 41.0y \;\;\; e = 0.79  \;\;\; a = 0\farcs87  
                                                   \nonumber    \\
   i = 62\fdg3   \;\;\;    \omega = 273\degr    \;\;\;  
                                              \Omega = 157\degr
\end{eqnarray}
for orbit ${\sc I}$;
\begin{eqnarray}
  P=98.7y  \;\;\; T = 38.3y \;\;\; e = 0.19  \;\;\; a = 0\farcs85  
                                                   \nonumber    \\
   i = 51\fdg1   \;\;\;    \omega = 243\degr    \;\;\;  
                                              \Omega = 98\degr
\end{eqnarray}
for orbit ${\sc II}$; and
\begin{eqnarray}
  P=4423.8y  \;\;\; T = 33.2y \;\;\; e = 0.95  \;\;\; a = 15\farcs2  
                                                   \nonumber    \\
   i = 73\fdg9   \;\;\;    \omega  = 207\degr    \;\;\;  
                                              \Omega  = 110\degr
\end{eqnarray}
for orbit ${\sc III}$.

\begin{figure}
\vspace{8.2cm}
\includegraphics{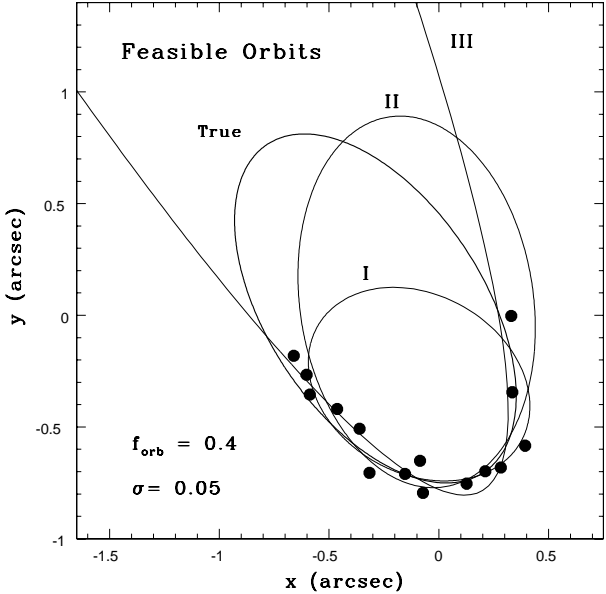}
\caption{Feasible orbits. Orbits ${\sc I,II,III}$ fit the  
measurements
(filled circles) with an acceptable $\chi^{2}$. The bold curve is the
true orbit with elements given in Eq.(8).}
\end{figure}

Fig.2 illustrates and confirms Aitken's warning (Sect. 1) that when 
$f_{orb} < 0.5$ several orbits will fit the data. Similar plots for real
binaries are given by Schaefer et al. (2006).

Orbit {\sc III} merits further comment. This orbit
is nearly parabolic and has its periastron during the observing campaign. Thus,
the observations were taken in time interval $(0,40y)$ and 
periastron occurs at $T = 33.2y$.
If this were the true orbit with $P = 4423.8y$, we would count ourselves
fortunate to catch a periastron passage in such a short campaign. 
This is a generic feature of fitting orbits to short observed arcs: the orbit
is not constrained when not observed and so can balloon out to large
angular separations.
This family of feasible orbits violate the
Copernican Principle since the observer is inferred to be at a special epoch.
Although this possibly justifies their exclusion, these orbits are 
retained in the calculation of posterior means.

\subsection{Eggen's effect}

To investigate Eggen's empirical discovery that even for poor orbits the ratio 
$a^{3}/P^{2}$
is relatively reliable, the ensemble of feasible orbits plotted
in Fig.1 is now projected onto the $(\log P, \log a)$ plane - Fig.3. From this
plot, we see 
a high density of points along the line   
\begin{equation}
   \log \: \frac{a}{a_{*}} =  \frac{2}{3} \: \log \: \frac{P}{P_{*}}   
\end{equation}
Thus, despite ranging in period from $40$ to $5000y$, the values of 
$a^{3}/P^{2}$ are relatively concentrated. This effect becomes even more
striking with improved campaigns having $f_{orb} = 0.5$ and $0.6$ - see
Fig.3.

These experiments indicate that with the chosen $(N,\sigma)$, the feasible
orbits are:\\
 1) closely confined to $(a_{*},P_{*})$ when $f_{orb} \ga 0.6$; \\
 2) dispersed widely in $P$ but narrowly in $a^{3}/P^{2}$ \\
when $0.5 \ga f_{orb} \ga 0.4$; \\
 3) dispersed widely in $P$ and increasingly so in $a^{3}/P^{2}$ 
when $f_{orb} \la 0.4$. \\   

These results provide strong theoretical support for Eggen's discovery and for
his use of the derived masses in studies of the mass-luminosity relation.
This confirmation derives from investigating the domain of high
likelihood in $(\log P,e,\tau)$-space. Since this is the domain in which
high weight is assigned to $Q$ when its posterior mean is calculated, 
Eggen's effect is evidently incorporated when this
formalism is used to estimate the binary's mass.
Accordingly, Fig.3 implies that we can anticipate
useful results in this standard case provided that $f_{orb} \ga 0.4$.    

The explanation of Eggen's effect is that as soon as 
secondary's relative motion reveals a significant departure from 
rectilinear motion ($f_{orb} \ll 1$) then acceleration has been detected 
and this is
$\propto ({\cal M}_{1} +  {\cal M}_{2}) \propto a^{3}/P^{2}$.
As $f_{orb}$ increases, the uncertainty due to orientation factors decreases
and so $a^{3}/P^{2}$ is well-determined before the orbit is complete
- see also Heintz (1978) and Schaefer et al. (2006).

\begin{figure}
\vspace{8.2cm}
\includegraphics{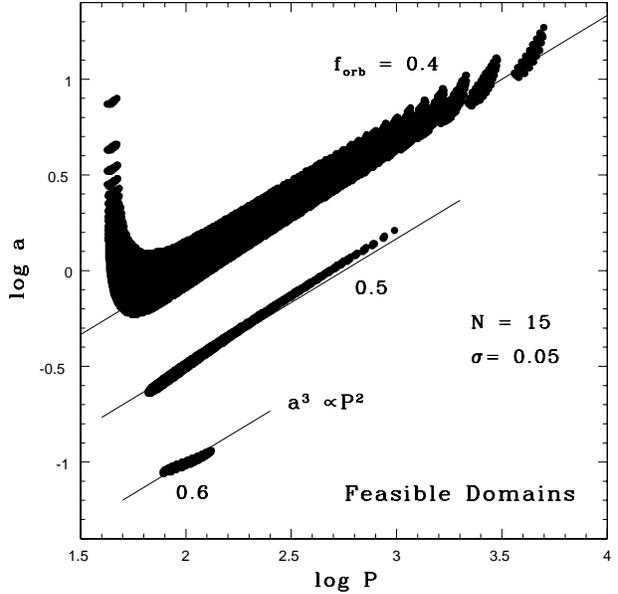}
\caption{Eggen's effect. Feasible domains ${\cal D}$ in the $(\log P, \log a)$
plane. Results for three campaigns with 
$f_{orb} = 0.4, 0.5, 0.6$ are shown, each with $N = 15, \sigma = 0\farcs05$.
The $f_{orb} = 0.5$ and $0.6$ domains are displaced downwards by 0.5 and 
1.0dex, respectively. The plotted lines have slope 2/3 and pass through
the points $(2.0,0.0), (2.0,-0.5), (2.0,-1.0)$, each corresponding
to $(\log P_{*}, \log a_{*})$.}
\end{figure}

In addition to the linear trends, Fig.3 shows a dramatic increase in the 
volume of ${\cal D}$ as $f_{orb}$ decreases from $0.6$ to $0.4$, supporting 
Aitken's warning about the onset of multiple acceptable orbits. Note that 
this increase in volume is a generic effect  
that is not sensitive to the choice of $0.05$ in Eq.(12).

This section has shown that the insights of Aitken and Eggen from
50-100 years ago find their modern explanation in how the
topology of the likelihood function ${\cal L}^{\dag}(\phi|D)$ responds
to changes in the observational data $D$.

\section{Mass estimates}

The analysis of Sects.2 and 3 is now applied to simulated partial orbits
for the standard visual binary. Results with variations of $e$ and ${\rm i}$
are reported in Appendix C.

\subsection{Mass estimator}

At the grid point $(i,j,k)$, the $\phi$ elements are $(P_{i},e_{j},\tau_{k})$, 
and the $\hat{\psi}$ elements are the least-squares values
$(\hat{A},\hat{B},\hat{F},\hat{G})$.
From the latter, the remaining Campbell elements
$(\hat{a},\hat{{\rm i}},\hat{\omega},\hat{\Omega})$ are computed as 
described in Sect.A.4, 
and we denote this semi-major axis as $\hat{a}_{ijk}$.  

Now if ${\cal M}_{*}$ is the true total mass of the model binary with
orbit defined by Eq.(8), the inferred mass with the orbit 
corresponding to the point $(i,j,k)$ is  
\begin{equation}
   {\cal M}_{ijk} = {\cal M}_{*} \left (\frac{\hat{a}_{ijk}}{a_{*}}  \right)^{3}    
                               \left(\frac{P_{i}}{P_{*}} \right)^{-2}
\end{equation}
The natural choice for the mass variable to be substituted in Eq.(7) 
is therefore  ${\cal M}_{ijk}$. However, numerical experiments demonstrate 
that $<\! \log {\cal M} \!>$  is more accurate than $\log <\!{\cal M} \!>$. 
Accordingly, we take
\begin{equation}
  Q_{ijk}^{\dag} = \log \frac{ {\cal M}_{ijk}}{ {\cal M}_{*}}
          = 3 \: \log \frac{\hat{a}_{ijk}}{a_{*}} - 2 \: \log \frac{P_{i}}{P_{*}}
\end{equation}
Exploiting the freedom to set ${\cal M}_{*} = 1$, we now 
substitute Eq.(18) into Eq.(7) to obtain the posterior mean 
$<\!\log {\cal M}\!>$,
where ${\cal M}$ is the ratio of the inferred to the exact mass. Thus, if
the exact mass is recovered, then $<\!\log {\cal M}\!> = 0$.

\subsection{Numerical experiments}

In these experiments, the true orbit is again given by Eq.(8). In the first
sequence of campaigns, $N = 15$ and $\sigma = 0\farcs05$ as in Sect.4,
but now $f_{orb}$ is treated as a continuous variable, thus investigating
the deterioration of mass estimates as $f_{orb} \rightarrow 0$.
Note that when $f_{orb}$ changes so does the seed for the random number 
generator. The pattern of gaussian variates in Eq.(11) is therefore never
repeated.

With steps of $0.01$ in $f_{orb}$, the values of  $<\!\log {\cal M}\!>$   
are plotted in Fig.4. We see that for $f_{orb} \ga 0.40$, the mass
estimates scatter about the exact value ${\cal M} = 1$ with errors 
$< 0.1$dex. However, for $f_{orb} \la 0.40$, the scatter increases sharply,
with the error first exceeding 0.1dex at $f_{orb} =  0.38$. These results
are consistent with the implications of Fig.3. 

Accuracy should improve with more data or data of higher 
precision - i.e, with decreasing precision parameter $\eta = \sigma/\sqrt{N}$.
This is investigated by reducing $\sigma$ from $0\farcs05$ to  $0\farcs005$
while keeping $N = 15$. This second sequence is plotted in Fig.4 as
filled circles. As expected, the quality of the estimates dramatically
inproves: the errors are $< 0.012$ dex for $f_{orb} \geq 0.36$, but 
increase sharply thereafter. Note that decreasing $\eta$ by a factor of 10 has only slightly
extended the domain of reliable solutions - the  error first exceeds 0.1dex at 
$f_{orb} = 0.34$. 

These experiments demonstrate that 
$<\!\log {\cal M}\!>$ 
provides a seemingly unbiased estimate of a visual binary's total mass 
even when standard orbit fitting 
generates multiple acceptable solutions with a large range of orbital periods.  

\begin{figure}
\vspace{8.2cm}
\includegraphics{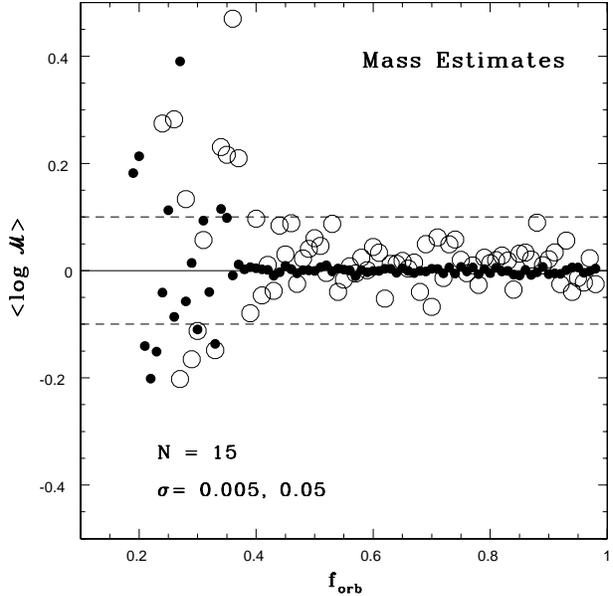}
\caption{Mass estimator. The posterior mean of $\log {\cal M}$ is plotted
against $f_{orb}$. The campaigns with $N = 15, \sigma = 0\farcs05$
are plotted as open circles; those with  $N = 15, \sigma = 0\farcs005$
as filled circles. The dashed lines indicate errors of $\pm 0.1$dex.}
\end{figure}

\subsection{Posterior pdf of $\log {\cal M}$}

For a real binary with an incomplete orbit, $f_{orb}$ is unknown.
Moreover, perhaps 
counterintuitively,
it cannot even be reliably estimated. Fig.2 and Eqs. (13)-(15) 
illustrate this:
for orbit {\sc I}, $f_{orb} = 0.85$; for orbit {\sc II}, $f_{orb} = 0.41$;
and for orbit {\sc III}, $f_{orb} = 0.009$, as against the true value $0.4$. 
It follows that an investigator cannot judge the accuracy of
$<\!\log {\cal M}\!>$ from an estimate of $f_{orb}$. Instead, the
posterior pdf of $\log {\cal M}$ should be derived and credible
intervals computed.

Inspection of Eq.(7) shows that the implied approximation to the
posterior pdf is
\begin{equation}
  \Theta(Q|D) = \Sigma_{ijk} W_{ijk} \delta(Q-Q_{ijk}^{\dag}) 
\end{equation}
where $\delta$ is the delta function, and 
\begin{equation}
  W_{ijk}  =   {\cal L}_{ijk}^{\dag} / 
                      \: \Sigma_{ijk} \: {\cal L}_{ijk}^{\dag}
\end{equation}
From these formulae, we immediately find
\begin{equation}
  \int \Theta \: dQ =1  \;\; and \;\;
               \int Q \: \Theta \: dQ = <\!Q\!>
\end{equation}
where $<\!Q\!>$ is the approximation given by Eq.(7).

A histogram representation of the pdf $\Theta(\log {\cal M}|D)$ is obtained by 
convolving Eq.(19) with a top-hat function. Examples are plotted in Fig.5
for $f_{orb} = 0.4$ and $0.5$. These show the loss of precision
with decreasing $f_{orb}$ that was expected from Fig.3 and seen in
Fig.4. 

The equivalent of $1 \sigma$ gaussian error bars can be computed from these 
pdf's by finding the equal area tails that together comprise $31.7\%$ of the
probability. Thus, we find 
$<\!\log {\cal M}\!> = 0.012^{+0.019}_{-0.019}$ for  $f_{orb} = 0.5$ and
$ = 0.068^{+0.077}_{-0.080}$ for $f_{orb} = 0.4$.

Because the pdf $\Theta(\log {\cal M}|D)$ is approximately gaussian - 
see Fig.5 - the above error bars can with sufficient
accuracy be replaced by the standard deviation of
the pdf, given by   
\begin{equation}
 s^{2}_{Q} = \Sigma_{ijk}  W_{ijk} (Q_{ijk}^{\dag} - <\!Q\!>)^{2}  
\end{equation}
This formula gives $s_{\log {\cal M}}   =0.019$ and $0.085$ for $f_{orb} = 0.5$ 
and $0.4$, respectively. 
Evidently, the dramatic loss in precision when the orbital coverage is 
insufficient is apparent from $s_{\log {\cal M}}$ even without knowing $f_{orb}$.

\begin{figure}
\vspace{8.2cm}
\includegraphics{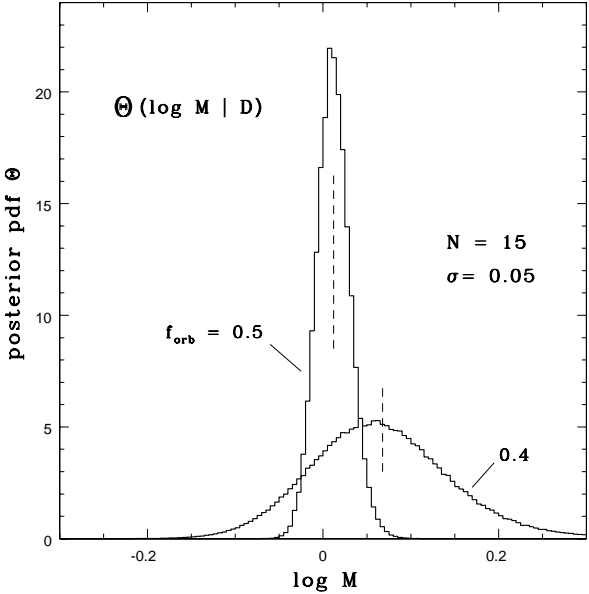}
\caption{Posterior pdf's $\Theta(\log {\cal M}|D)$ for $f_{orb} = 0.4$ and
$0.5$ both with $N = 15,\: \sigma = 0\farcs05$. The vertical dashed lines
are the posterior means  $<\!\log {\cal M}\!>$.}
\end{figure}

\subsection{Bias}

For $f_{orb} \ga 0.4$, Fig.4 shows no evidence that the estimator 
$<\! \log {\cal M} \!>$ is biased. 
But for $f_{orb} \la 0.4$, any bias is obscured by the dramatic 
increase in scatter.
Accordingly, since quantifying
the onset of bias 
is of interest in understanding the limits of this Bayesian approach, 
further calculations are now reported.

In order to beat down the noise, the campaigns with 
$N=15, \: \sigma = 0\farcs05$
are repeated $n = 100$ times for selected values of $f_{orb}$. The average
and variance
\begin{equation}
  \bar{Q} = \frac{1}{n} \Sigma_{i} <\!Q\!>_{i}  \;\;\;  
                  s^{2} =\frac{1}{n-1} \Sigma_{i} (<\!Q\!>_{i} -\bar{Q})^{2}  
\end{equation}
of the $<\:Q\:>_{i}$ are computed, so that the resulting estimate  is 
$\bar{Q} \pm s/\sqrt{n}$, and this indicates bias if it differs significantly 
from the exact value of $Q$.

Fig.6 presents evidence of bias for the estimator $<\!\log ({\cal M})\!>$. 
The average values  of $<\!\log ({\cal M})\!>$ and their 
standard deviations from 100 independent realisations of the orbits are 
plotted for $f_{orb} = 0.30(0.05)0.95$. There is no evidence of bias for
$f_{orb} \geq 0.6$, but strong evidence for $f_{orb} \leq 0.4$ where
the displacements are $ > 3.5 s$. Similar plots are given in Appendix B
for the posterior means of the Campbell elements.  

The onset of bias coincides with the onset of scatter (see Fig.4)
at which point accurate mass determinations are no longer possible.
Nevertheless, from a technical standpoint, it is of interest to understand
the origin of bias. Comparison of Figs. 2,3, and 6 shows that the onset of bias
corresponds to the dramatic increase in the volume of ${\cal D}$, 
and this implies that the choice of priors becomes increasingly 
important. Accordingly, since the expansion of ${\cal D}$ brings in orbits of 
high eccentricity (Fig.2), we might seek to penalize such orbits
via the prior on $e$. But this would be an {\em ad hoc} fix
for this {\em particular} model binary. A fundamental solution probably
invokes the Copernican Principle (Sect.4.1) to reduce the weight    
assigned to highly eccentric orbits.

\begin{figure}
\vspace{8.2cm}
\includegraphics{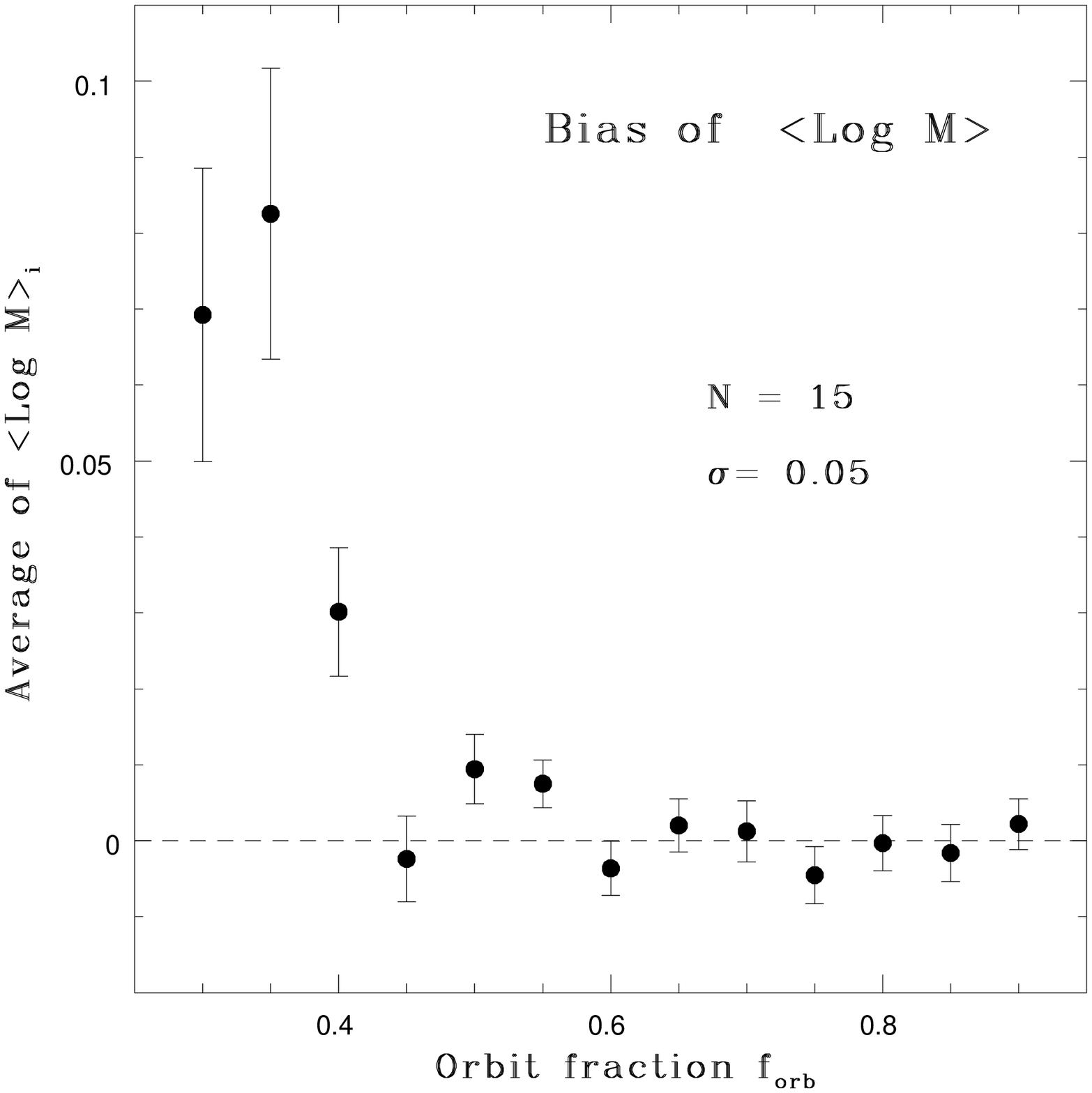}
\caption{Bias of the posterior mean $<\!\log ({\cal M})\!>$.
The averages and standard errors from Eq.(23) plotted against $f_{orb}$
with $N=15, \sigma = 0\farcs05$. The dashed line is the exact value.}
\end{figure}

\subsection{Coverage fractions}

When the estimate of a parameter derived by least-squares analysis is reported
as $\bar{Q} \pm \sigma_{Q}$,
we automatically interpret this as an assertion that there is a $68.3 \%$ 
probability that the true answer lies in the interval 
$(\bar{Q} - \sigma_{Q}, \bar{Q} + \sigma_{Q})$. 
Implicit in this interpretation, however, are the following assumptions:
1) normally-distributed measurement errors; 
2) the validity of the linearization used in deriving the normal equations; 
and 3) that the number of data points $N$ greatly 
exceeds $M$, the number of parameters estimated. If one or more these 
assumptions does
not hold then the above interval's {\em coverage fraction} will differ 
from $0.683$.
   
In Bayesian statistics, credibility intervals derived from the posterior
pdf replace the confidence intervals of frequentist statistics, and the
different meanings of these intervals has been topic of much discussion
in the statistical literature over many years. One approach to reconciling
this aspect of the Bayesian-frequentist debate is that of calibrated
Bayes (e.g., Dawid 1982, Little 2006): A credible interval is 
{\em well-calibrated} if the enclosed probability matches the fraction of times
that it contains the true value.

In order to test whether credible intervals calculated with this formalism
for visual binaries are well-calibrated, 1000 independent realizations
of two campaigns with  
$N=15, \sigma = 0\farcs05$ have been analysed, one with $f_{orb} = 0.60$,
the other with $f_{orb} = 0.95$. For each realization, we check if
$Q_{exact}$ is in the intervals    
$(<\! Q \!> - m s_{Q}, <\! Q \!> + m s_{Q})$ for $m = 1,2$, where 
$<\! Q \!> $ is given by Eq(7) and $s_{Q}$ by Eq.(22).

The results of this exercise are in Table 1. The first row gives the desired 
fraction ${\cal E}(f)$ - i.e., the fraction expected for an {\em ideal} 
frequentist analysis. The remaining rows give the results for the 
posterior means of the Campbell
elements and $\log {\cal M}$. 

Inspection of these results shows that
the coverage is close to ideal for the $\phi$ elements $P,e,\tau$. Accordingly,
their credible intervals are well-calibrated. However, for the remaining
elements and for $\log {\cal M}$, the error bars are too small by factors
of $1.1-2.1$ at $1 \sigma$ and $1.1-1.6$ at $2 \sigma$. 
These error bars could of course be calibrated from such 
simulations. However, these shortfalls occur for the elements that
were replaced by the Thiele-Innes constants and thus had the widths
of their posterior pdfs reduced by the delta function approximation in Eq.(3).
Further research based on the analytical formulae of Sect.A.3 may 
alleviate or eliminate this problem.

\begin{table}

\caption{Coverage fractions $f$ from $1000$ trials}

\label{table:1}

\centering

\begin{tabular}{c c c c}

\hline
\hline

 $Q$                 &  $f_{orb}$ &     $1\sigma$       &  $ 2\sigma$      \\

\hline                                                                  

                                                                         \\

 ${\cal E}(f)$     &           &   $0.683 \pm 0.015$  &    $0.954 \pm 0.007$ \\ 

                                                                         \\

 $<\! \log P \!>$  &  0.60     &   $0.680 \pm 0.015$  &  $0.945 \pm 0.007$ \\ 

                   &  0.95     &   $0.697 \pm 0.015$  & $0.962 \pm 0.006$  \\

 $<\! e \!>$       &  0.60     &   $0.656 \pm 0.015$  & $ 0.954 \pm 0.007$ \\ 

                   &  0.95     &   $0.670 \pm 0.015$  & $0.948 \pm 0.007$ \\

 $<\! \tau \!>$    &  0.60     &   $0.677 \pm 0.015$  & $ 0.945 \pm 0.007$ \\

                   &  0.95     &   $0.656 \pm 0.015$  & $0.957 \pm 0.006$  \\

                                                        \\

 $<\! \log a \!>$  &  0.60     &   $0.536 \pm 0.016$  & $0.858 \pm 0.011$ \\

                   &  0.95     &   $0.388 \pm 0.015$  & $0.709 \pm 0.014$  \\

 $<\! i \!>$       &  0.60     &   $0.386 \pm 0.015$  & $ 0.680 \pm 0.015$ \\

                   &  0.95     &   $0.491 \pm 0.016$  & $0.826 \pm 0.012$  \\

 $<\! \omega \!>$  &  0.60     & $0.607 \pm 0.015$    & $0.892 \pm 0.010$  \\

                   &  0.95     & $0.516 \pm 0.016$    & $0.831 \pm 0.012$  \\

 $< \! \Omega \!>$ &  0.60     & $0.398 \pm 0.015$   & $0.702 \pm 0.014$  \\

                   &  0.95     & $0.331 \pm 0.015$    & $0.611 \pm 0.015$  \\

                                                                         \\

 $<\! \log {\cal M} \!>$  & 0.60  & $0.371 \pm 0.015$  & $0.691 \pm 0.015$  \\

                          & 0.95  & $0.480 \pm 0.016$ & $0.786 \pm 0.013$    \\

                                                               \\

\cline{1-4}

\hline
\hline

\end{tabular}

\end{table}

\section{Conclusion}

The aim of this paper has been to apply Bayesian methods to a 
problem that dates back a century or more, namely
how to estimate the total mass of a visual binary that
has a measured parallax but an incomplete orbit. This problem has 
previously been
effectively treated  by Schaefer et al. (2006)
with a heuristic Monte Carlo search technique.
In contrast, the Bayesian approach has a firm theoretical 
foundation and is, moreover, now the preferred methodology  
in many areas of astronomy, 
notably in cosmology (e.g., Liddle 2009) and in the discovery and 
characterization of the orbits of exoplanets (e.g., Ford \& Gregory 2007). 

A further merit of Bayesian estimation  
for visual binaries is that only triple integrals need to 
be evaluated
when the quadrivariate normal distribution of the Thiele-Innes constants 
(Sect.A.3) is approximated by a delta function (Sect.2.2). The use of this 
approximation might be expected to introduce
systematic errors in the derived elements. However, the failure to
detect bias when $f_{orb} \ga 0.5$ (Sect.5.4, Appendix B) demonstrates that
such errors are not of practical concern.

In Sect.1, the possible application to binaries discovered by {\em Gaia}
was mentioned. However, the simulations reported in this paper relate
to ground-based data where {\em two} sky coordinates are measured.
In contrast, the scanning mode of {\em Gaia} will provide $\sim 70$
high precision {\em one} dimensional measurements of a secondary's
displacement (e.g., Pourbaix 2002; Lattanzi et al. 2000). 
Simulations of a model binary observed in this
manner are required to demonstrate the merits of this Bayesian approach
for {\em Gaia}.

\appendix  

\section{Thiele-Innes elements}

In this Appendix, a concise self-contained account of the Thiele-Innes elements
is presented - see also Wright \& Howard (2009) and Hartkopf et al. (1989). 
The explicit least-squares formulae for $A,B,F,G$ and for their variances and
covariances are new - Eqs.(A.7),(A.9),(A.10).

\subsection{Definitions}

The Thiele-Innes constants $A,B,F,G$ are defined in terms of four of the
Campbell elements. Specifically, 
\begin{eqnarray}
 A=a(+\cos \: \Omega \: \cos \: \omega - \sin \: \Omega \: \sin \: \omega \: 
                                   \cos \: i)  \nonumber \\
 B=a(+\sin \: \Omega \: \cos \: \omega + \cos \: \Omega \: \sin \: \omega \: 
                                   \cos \: i)  \nonumber \\
 F=a(-\cos \: \Omega \: \sin \: \omega - \sin \: \Omega \: \cos \: \omega \: 
                                   \cos \: i)  \nonumber \\
 G=a(-\sin \: \Omega \: \sin \: \omega + \cos \: \Omega \: \cos \: \omega \: 
                                   \cos \: i)
\end{eqnarray}
Given these constants, the coordinates $(x,y)$ of the 
secondary relative to the primary at time $t$ are
\begin{eqnarray}
  x(t)= A \: X(E) + F \: Y(E)             \nonumber    \\  
  y(t) = B \: X(E) + G \: Y(E) 
\end{eqnarray}
where
\begin{equation}
  X = \cos \: E - e    ,   \;\;   Y = \sqrt{1-e^{2}} \: \sin \: E
\end{equation}
and $E(t)$, the eccentric anomaly, is given by Kepler's equation
\begin{equation}
  \mu(t-T) = E - e \: \sin \: E  \;\;\; with  \;\;\; \mu = 2 \pi/P
\end{equation}

\subsection{Least-squares estimates}

On the assumption that $P,e,T$ are known, the 
determination of the four Thiele-Innes elements is a straightforward problem 
of fitting a {\em linear} model to the data.
If the observed position at $t_{n}$ is $(\tilde{x}_{n},\tilde{y}_{n})$, then
estimates for $A,B,F,G$ may be obtained by minimizing
\begin{equation}
  \chi^{2} = \frac{1}{\sigma^{2}} \Sigma_{n} w_{n} (x_{n}-\tilde{x}_{n})^{2}
   +\frac{1}{\sigma^{2}} \Sigma_{n} w_{n} (y_{n}-\tilde{y}_{n})^{2}
\end{equation}
where $(x_{n},y_{n})$ is the theoretical position given by Eq.(A.2).
Here $w_{n}$ is the weight of the $n$-th observation and $\sigma$ is
the standard error of a measurement of unit weight.

This least-squares problem can be solved without iteration since $x$ and
$y$ are linear in $A,B,F,G$. Moreover, the first term in
Eq.(A.5) depends only on the pair $(A,F)$ and the second term only on $(B,G)$.
Accordingly, the estimates for $A,F$ are obtained by minimizing the first term
in Eq.(A.5) and those for $B,G$ by minimizing the second term. Thus, the 
problem reduces to solving two independent pairs of linear equations, each 
with two unknowns. 

To simplify the resulting formulae, we define the following quantities 
\begin{eqnarray}
 a = \Sigma_{n} w_{n}  X_{n}^{2}  \;\;\; b = \Sigma_{n} w_{n} Y_{n}^{2}  \;\;\; 
                    c = \Sigma_{n} w_{n} X_{n} Y_{n} \nonumber \\
 r_{11} = \Sigma_{n} w_{n}  \tilde{x}_{n} X_{n}     \;\;\;\;\;
 r_{12} = \Sigma_{n} w_{n}  \tilde{x}_{n} Y_{n}                      \\
 r_{21} = \Sigma_{n} w_{n}  \tilde{y}_{n} X_{n}     \;\;\;\;\;
 r_{22} = \Sigma_{n} w_{n}  \tilde{y}_{n} Y_{n}   \nonumber 
\end{eqnarray}
The least-squares estimates for the Thiele-Innes constants are then 
\begin{eqnarray}
 \hat{A} = (b r_{11} - c r_{12})/\Delta     \;\;\;\;\;
 \hat{F} = (-c r_{11} + a r_{12})/\Delta          \nonumber   \\    
 \hat{B} = (b r_{21} - c r_{22})/\Delta     \;\;\;\;\;
 \hat{G} = (-c r_{21} + a r_{22})/\Delta           
\end{eqnarray}
where  $\Delta = ab - c^{2}$. 

The above analysis is readily generalized to allow
$\tilde{x}_{n},\tilde{y}_{n}$ to have different weights $w_{n}^{x},w_{n}^{y}$.
In the formulae for $(\hat{A},\hat{F})$, the quantities $a,b,c,r_{11},r_{12}$ are evaluated
with $w_{n} = w_{n}^{x}$. Correspondingly,   
in the formulae for $(\hat{B},\hat{G})$, the quantities $a,b,c,r_{21},r_{22}$ are evaluated
with $w_{n} = w_{n}^{y}$.   

\subsection{Covariance matrix}

The two sets of normal equations solved above for $(\hat{A},\hat{F})$ and
$(\hat{B},\hat{G})$ have the same matrix of coefficients and therefore the same inverse.
These are
\begin{equation}
 \vec{M} =   \left(\begin{array}{clcr}
       a & c \\
       c & b
     \end{array} \right)     \;\;\; and \;\;\;   
 \vec{M^{-1}} =   \left(\begin{array}{clcr}
      +b/\Delta & -c/\Delta \\
      -c/\Delta  & +a/\Delta
     \end{array} \right)   
\end{equation}
From the coefficients of the covariance matrix $\vec{M^{-1}}$,
we derive
\begin{equation}
 \sigma^{2}_{A,B} = \frac{b}{\Delta} \sigma^{2}  \;\;\;\;
    \sigma^{2}_{F,G} = \frac{a}{\Delta} \sigma^{2}
\end{equation}
and
\begin{equation}
 cov(A,F) =  cov(B,G) = -\frac{c}{\Delta} \sigma^{2} 
\end{equation}
The covariances of other pairings of $A,B,F,G$ are zero.
Note that these results apply for fixed $P,e,T$. 

The above analysis shows that $(A,F)$ have a bivariate
normal distribution centred on $(\hat{A}, \hat{F})$ with variances and 
covariance from Eqs. (A.9) and (A.10) - and correspondingly for the pair 
$(B,G)$.   
Accordingly,
the pdf $p(A,B,F,G|\phi)$ is a quadrivariate normal distribution centred on
$(\hat{A}, \hat{B}, \hat{F}, \hat{G}) = \hat{\psi}(\phi)$.    
This explicit derivation of $p(\psi|\phi)$ could be used to replace the 
delta function in Eq.(3).

\subsection{Campbell elements}

Having derived the least-squares values of $A,B,F,G$ from Eqs.(A.7),
we must invert the transformation given by Eq.(A.1).
The angle $\omega+\Omega$ is the solution of the equation
\begin{equation}
  \omega+\Omega = \arctan \left(\frac{B-F}{A+G} \right)
\end{equation}
such that $\sin(\omega+\Omega)$ has the same sign as $B-F$. Similarly,
$\omega-\Omega$ is the solution of the equation
\begin{equation}
  \omega-\Omega = \arctan \left(\frac{-B-F}{A-G} \right)
\end{equation}
such that $\sin(\omega-\Omega)$ has the same sign as $-B-F$. 
This pair of equations can be solved for $\omega$ and $\Omega$. 
But $\Omega$ may violate the convention that $\Omega \in (0,\pi)$.
Accordingly, if  $\Omega < 0$, we set  $\Omega = \Omega + \pi$
and  $\omega = \omega + \pi$. But if $\Omega > \pi$, we set 
$\Omega = \Omega - \pi$ and  $\omega = \omega - \pi$.

If we now define
\begin{equation}
  q_{1} = \frac{A+G}{\cos(\omega+\Omega)}  \;\; and \;\;
                           q_{2} = \frac{A-G}{\cos(\omega-\Omega)}
\end{equation}
then the inclination is
\begin{equation}
  i = 2 \: \arctan (\sqrt{q_{2}/q_{1}} \:)
\end{equation}
and the semi-major axis is
\begin{equation}
  a= (q_{1}+q_{2})/2 
\end{equation}

\section{Bias of posterior means}

The numerical search for bias 
reported in Sect.5.4 also included  
the Campbell elements.
The results are plotted in Figs.B1-B7. The behaviour revealed by these plots
is broadly similar to that seen in Fig.6 except that for some elements bias is
evident at somewhat larger $f_{orb}$. 
Thus biases $> 3s$ occur at $f_{orb} = 0.5$ for $<\! \log P \!>,<\! \log a \!>, <\! \tau \!>$ and $<\!e\!>$.   

\begin{figure}
\vspace{8.2cm}
\includegraphics{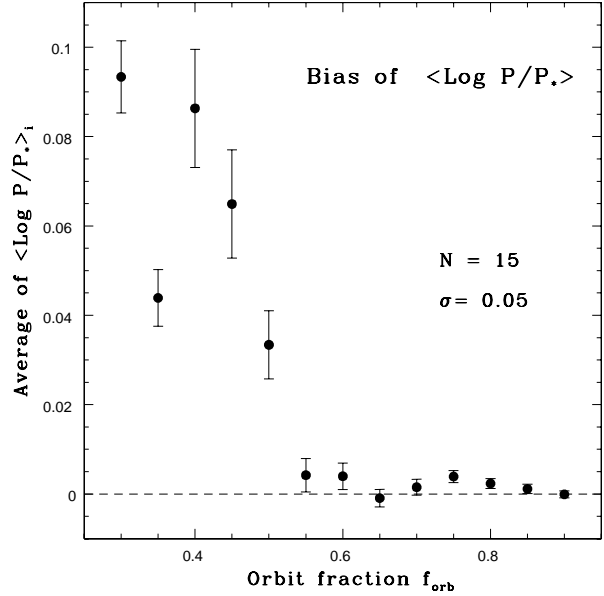}
\caption{Bias of the posterior mean $<\!\log (P/P_{*})\!>$.
The averages and standard errors from Eq.(23) plotted against $f_{orb}$
with $N=15, \sigma = 0\farcs05$. The dashed line is the exact value.}
\end{figure}
\begin{figure}
\vspace{8.2cm}
\includegraphics{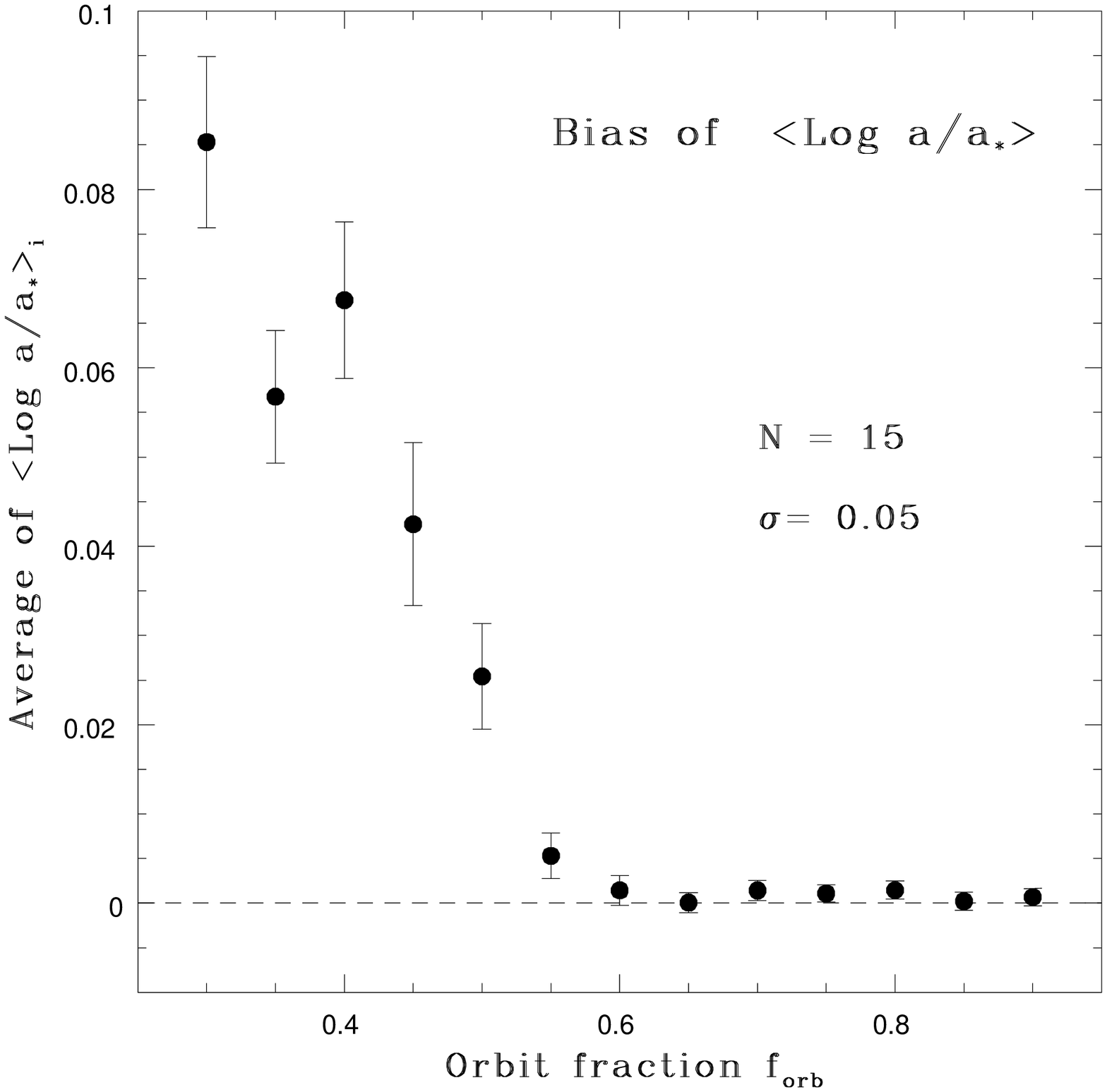}
\caption{Bias of the posterior mean $<\!\log (a/a_{*})\!>$.
The averages and standard errors from Eq.(23) plotted against $f_{orb}$
with $N=15, \sigma = 0\farcs05$. The dashed line is the exact value.}
\end{figure}
\begin{figure}
\vspace{8.2cm}
\includegraphics{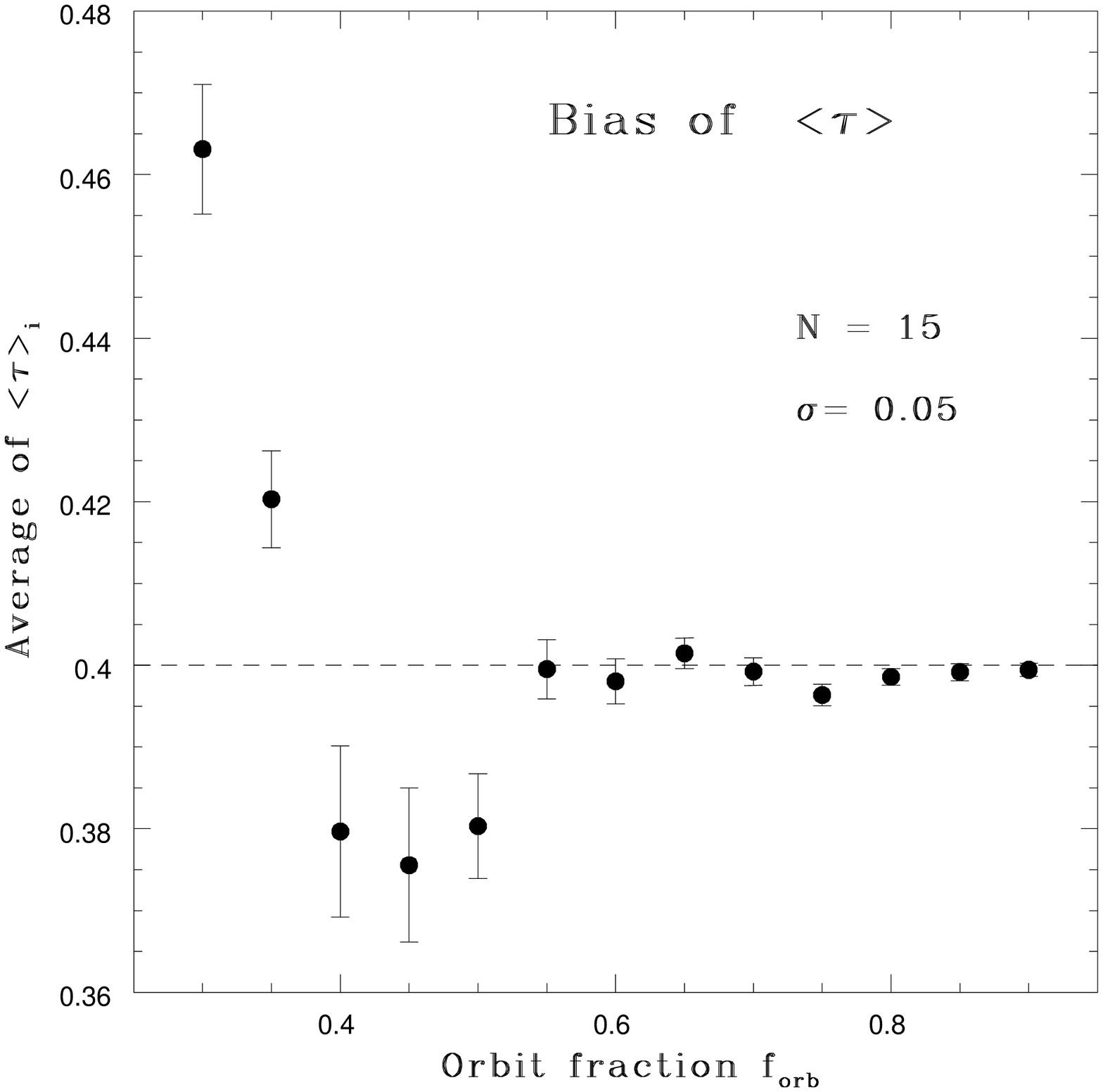}
\caption{Bias of the posterior mean $<\!\tau\!>$.
The averages and standard errors from Eq.(23) plotted against $f_{orb}$
with $N=15, \sigma = 0\farcs05$. The dashed line is the exact value.}
\end{figure}
\begin{figure}
\vspace{8.2cm}
\includegraphics{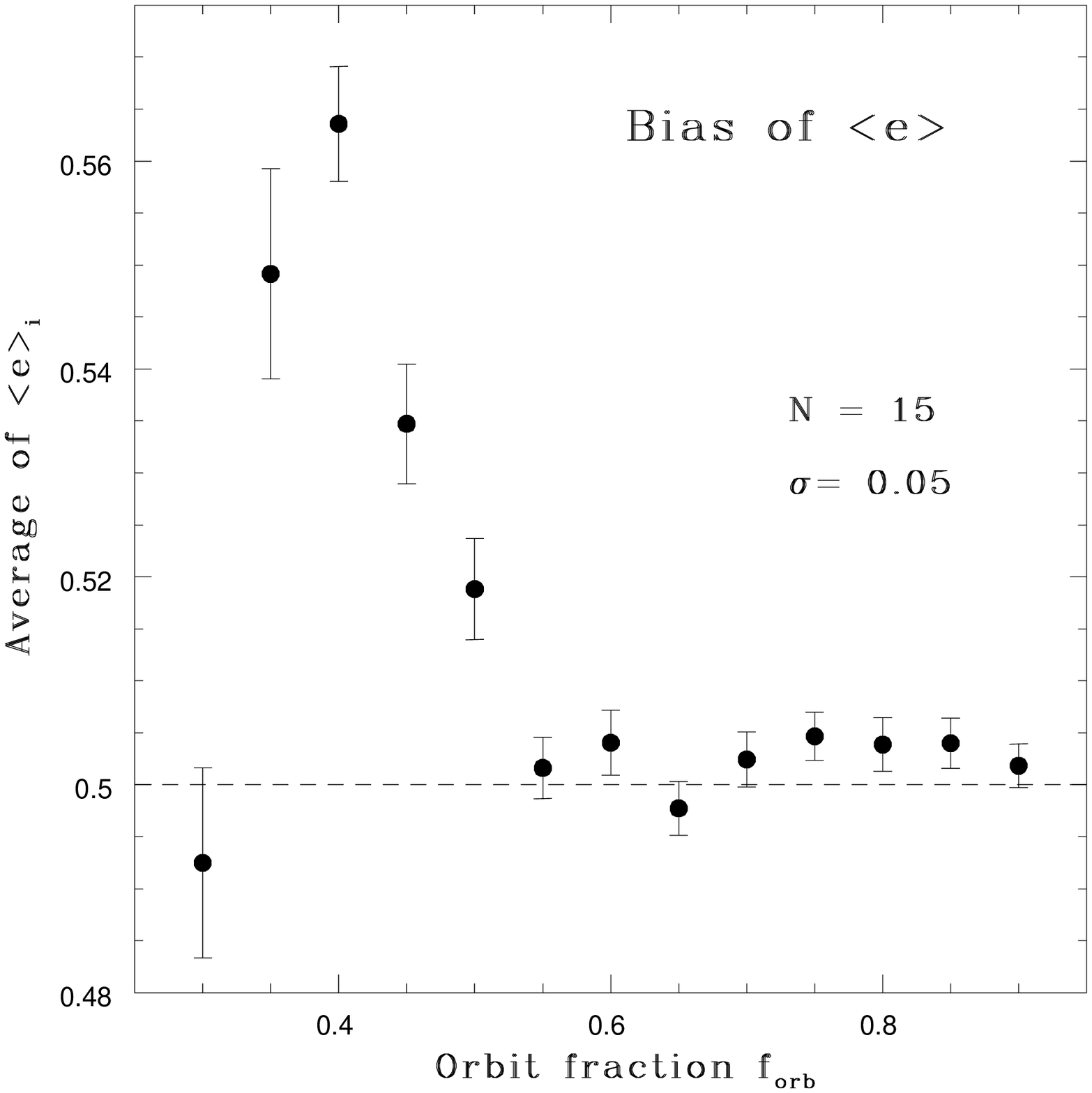}
\caption{Bias of the posterior mean $<\!e\!>$.
The averages and standard errors from Eq.(23) plotted against $f_{orb}$
with $N=15, \sigma = 0\farcs05$. The dashed line is the exact value.}
\end{figure}
\begin{figure}
\vspace{8.2cm}
\includegraphics{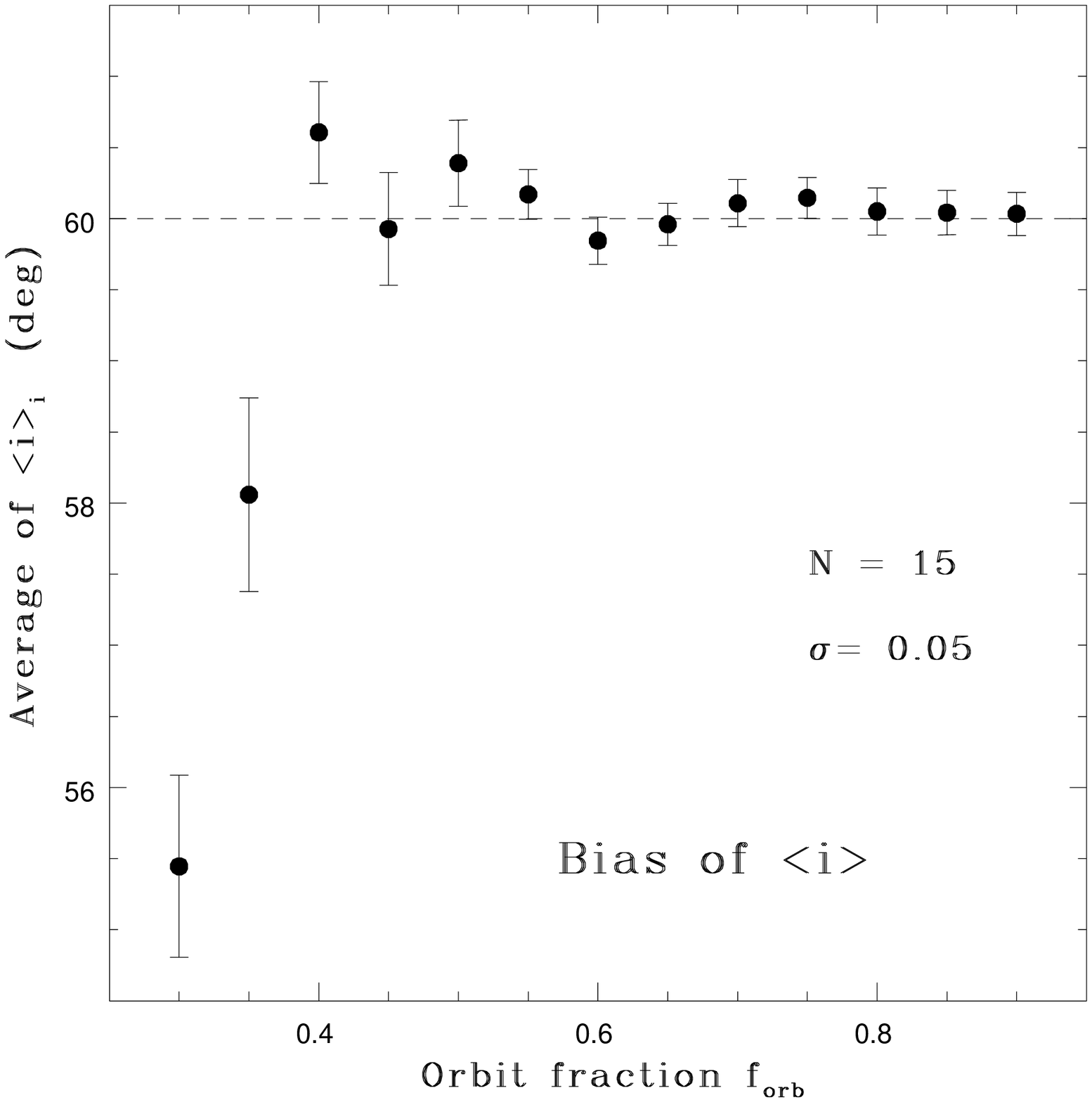}
\caption{Bias of the posterior mean $<\!\rm i\!>$.
The averages and standard errors from Eq.(23) plotted against $f_{orb}$
with $N=15, \sigma = 0\farcs05$. The dashed line is the exact value.}
\end{figure}
\begin{figure}
\vspace{8.2cm}
\includegraphics{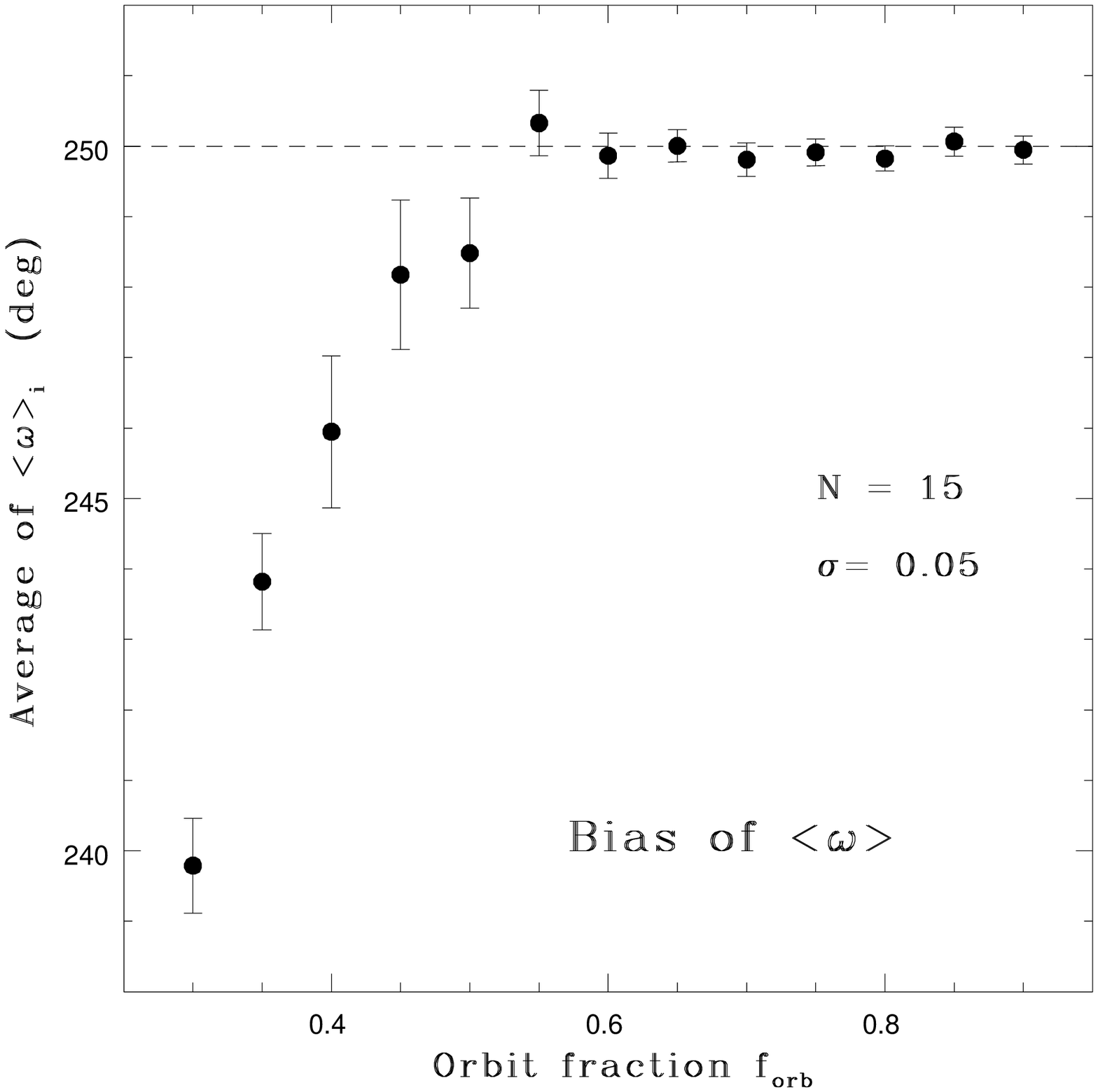}
\caption{Bias of the posterior mean $<\!\omega\!>$.
The averages and standard errors from Eq.(23) plotted against $f_{orb}$
with $N=15, \sigma = 0\farcs05$. The dashed line is the exact value.}
\end{figure}
\begin{figure}
\vspace{8.2cm}
\includegraphics{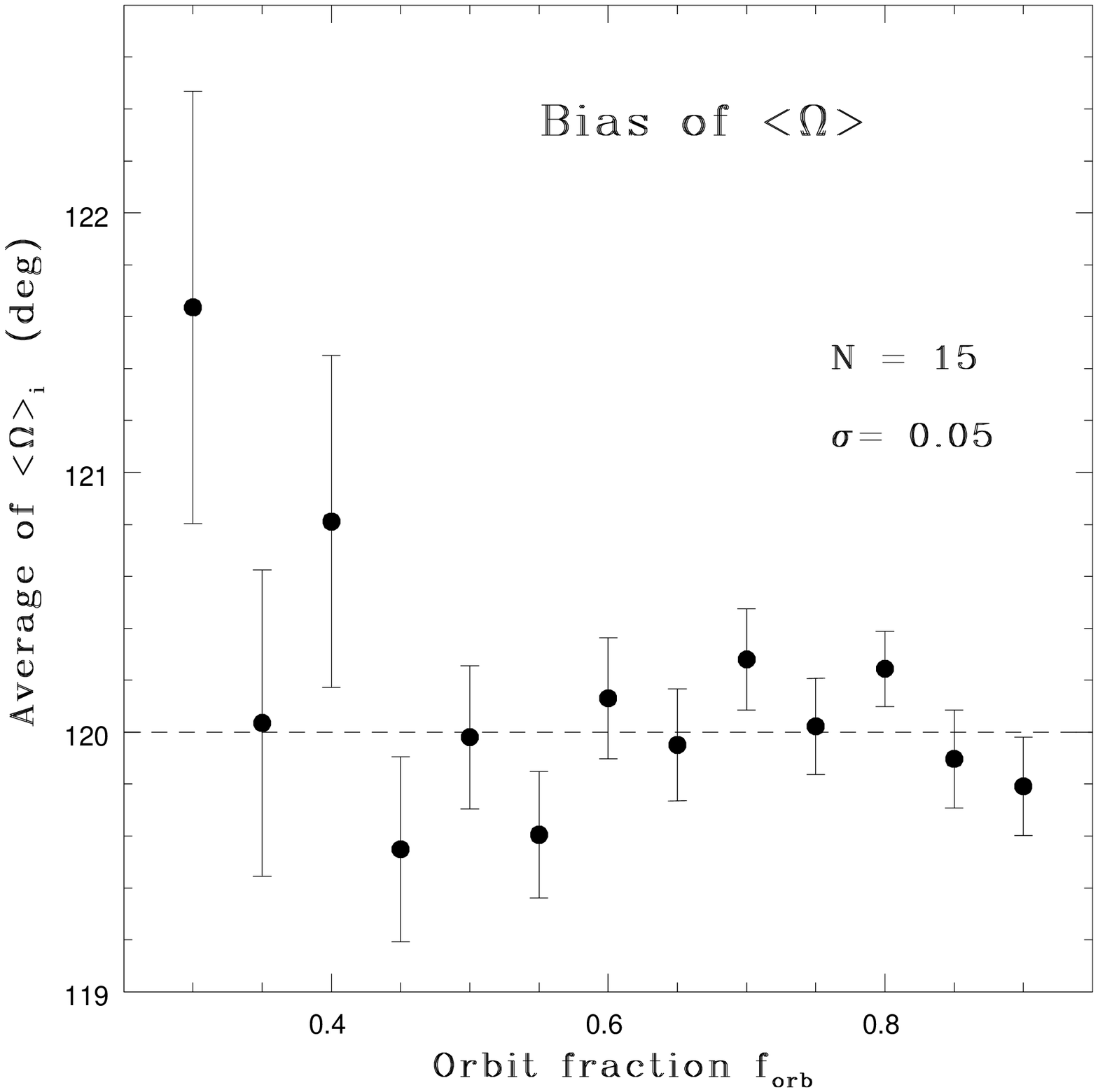}
\caption{Bias of the posterior mean $<\!\Omega\!>$.
The averages and standard errors from Eq.(23) plotted against $f_{orb}$
with $N=15, \sigma = 0\farcs05$. The dashed line is the exact value.}
\end{figure}

\section{Variation of parameters}

In this Appendix, posterior means are computed when selected orbital 
parameters are varied from those of the standard model - Eq.(8).
Specifically, in one sequence 
$e = 0.00(0.19)0.95$, and in a second sequence 
${\rm i} = 0\degr(18\degr)90\degr$, with, in each case, the other elements as
in Eq.(8). For both sequences, the campaign parameters are 
$f_{orb} = 0.7, N = 15, \sigma = 0\farcs05$, and each orbit is computed
with a different random number seed.
The orbits are poor and 
incomplete but 
$f_{orb}$ is in the domain $(f_{orb} \ga 0.6)$ where bias due to priors is
negligible - Sect.5.4 and Appendix B.

Results for the $e$-sequence are in Table C.1. Aspects to note are as follows:
The solution for $e = 0$
has large error bars for $<\! \omega \!>$ and $<\! \tau \!>$. These arise 
because
$\omega$ and $\tau = T/P$ become meaningless and indeterminate as
$e \rightarrow 0$ - i.e., for a circular orbit. Also since $e$ is
non-negative,  $<e>$ is positively-biased when the true value is 
$e = 0$. Despite these issues, an accurate mass is obtained.

At the other extreme of  highly elliptical orbits, Table C.1 reveals large 
error bars for all elements except $<\! e \!>$.

Results for the $i$-sequence are in Table C.2. Aspects to note are as follows:
The solution ${\rm i} = 0\degr$  - i.e., a face-on orbit - has large error
bars for $\omega$ and $\Omega$. From Eqs. (A.1), we see that when 
${\rm i} = 0\degr$, $A = G = a \: cos (\omega + \Omega)$ and 
$B = -F = a \: sin (\omega + \Omega)$. Accordingly, for an exactly face-on 
orbit,
the difference $\omega - \Omega$ is indeterminate, and the error bars on 
$\omega$ and $\Omega$ reflect this. Nevertheless, the mass determination
is accurate.

As was the case for $<\! e \! >$, the quantity $<\! {\rm i} \!>$ is 
postively-biased
since ${\rm i}$ is non-negative. The expected magnitude of this bias can
be computed independently of the Bayesian code. If we assume that 
$P,e,\tau$ have the values given in Eq.(8), then the least-squares fit
to a synthetic orbit with ${\rm i} = 0$ is derived with Eqs. (A.7) and then 
the corresponding Campbell elements $a,{\rm i},\omega,\Omega$ 
obtained with Eqs.(A.11)-(A.15). With 
the chosen campaign parameters, $n = 1000$ repetitions of this calculation  
gives the average value $\bar{{\rm i}} = 13\fdg9$, consistent with 
$14\fdg4 \pm 4\fdg0$ in Table C.2. When $\sigma$ is reduced to $0\farcs005$
as in Sect.5.2,  $\bar{{\rm i}} = 4\fdg4$. Thus, as expected, the bias of
${\rm i}$ decreases with improved data.
 
A further point of note in the $i$-sequence is that the standard error
of $<\! \Omega \!>$ is much smaller than that of $<\! \omega \!>$ when 
${\rm i} = 90\degr$ - i.e., an edge-on orbit. Although the error bars
of these elements are somewhat underestimated (Sect.5.5), this is a
real effect. From Eqs. (A.1) and (A,2), it follows that the exact
edge-on orbit obeys the equation $y/x = tan \: \Omega$, so that
$\Omega$ is determined by the slope of the straight line fitted
to $\tilde{x}_{n},\tilde{y}_{n}$.  The Bayesian code does not 
formally make such a fit, but this mathematical aspect is implicit.

\begin{table*}
\begin{minipage}{80mm}

\caption{Sequence with varying eccentricity $e$}

\label{table:C.1}

\centering

\begin{tabular}{c c c c c c c}

\hline
\hline

 $Q$            &  $e = 0$ &     0.19       &  0.38  & 0.57 & 0.76 & 0.95    \\

\hline                                                                  
                                                                       \\

 $<\! \log P \!>$  & $1.994 \pm 0.013$   & $2.024 \pm 0.019$    
            &  $2.014 \pm 0.017$  &  $2.011 \pm 0.018$ & $1.997 \pm 0.019$
            &  $1.976 \pm 0.038$   \\

 $<\! e \!>$       &  $0.022 \pm 0.018$     &  $0.237 \pm 0.029$  
  & $0.402 \pm 0.025$    & $0.570 \pm 0.024$ & $0.783 \pm 0.025$
  & $0.941 \pm 0.026$  \\

 $<\! \tau \!>$    & $0.524 \pm 0.249$     &   $0.376 \pm 0.020$    
   & $0.382 \pm 0.016$  & $0.392 \pm 0.016$  &  $0.405 \pm 0.018$
   &  $0.424 \pm 0.036$  \\

                                                        \\

 $<\! \log a \!>$  & $-0.002 \pm 0.004$      &   $0.003 \pm 0.005$      
            & $-0.005 \pm 0.004$ & $-0.014 \pm 0.007$ & $0.036 \pm 0.018$
            &  $0.011 \pm 0.109$  \\

 $<\! {\rm i} \!>$       &  $59\fdg7 \pm 0\fdg6$      &  $59\fdg1 \pm 0\fdg3$  
  & $60\fdg0 \pm 0\fdg6$ & $57\fdg7 \pm 1\fdg1$ & $63\fdg8 \pm 1\fdg6$
  &  $62\fdg6 \pm 5\fdg5$   \\

 $<\! \omega \!>$  & $225\fdg6 \pm 100\fdg7$     &  $246\fdg9 \pm 4\fdg0$ 
    & $245\fdg3 \pm 2\fdg2$  & $247\fdg5 \pm 1\fdg8$ & $253\fdg4 \pm 1\fdg6$ 
    &  $247\fdg0 \pm 7\fdg3$ \\

 $< \! \Omega \!>$ &  $122\fdg1 \pm 1\fdg1$      &  $120\fdg9 \pm 1\fdg3$  
   & $121\fdg8 \pm 1\fdg1$  & $121\fdg2 \pm 1\fdg3$ & $117\fdg8 \pm 1\fdg3$ 
   &  $125\fdg4 \pm 5\fdg5$  \\

                                                                         \\

 $<\! \log {\cal M} \!>$  & $0.011 \pm 0.026$   & $-0.048 \pm 0.037$  
            & $-0.028 \pm 0.034$  & $-0.022 \pm 0.036$ & $0.007 \pm 0.039$
            & $0.059 \pm 0.124$ \\

                                                                  \\

\cline{1-7}

\hline
\hline

\end{tabular}

\end{minipage}
\end{table*}

\begin{table*}
\begin{minipage}{80mm}

\caption{Sequence with varying inclination ${\rm i}$}

\label{table:C.2}

\centering

\begin{tabular}{c c c c c c c}

\hline
\hline

 $Q$   &  ${\rm i} = 0\degr$ &  $18\degr$   &  $36\degr$ &  $54\degr$ 
                                            & $72\degr$ & $90\degr$   \\

\hline                                                                  
                                                                       \\

 $<\! \log P \!>$  & $1.998 \pm 0.013$   & $1.999 \pm 0.013$    
            &  $2.043 \pm 0.018$  &  $1.994 \pm 0.014$ 
            & $2.011 \pm 0.017$   & $1.965 \pm 0.015$  \\

 $<\! e \!>$       &  $0.504 \pm 0.016$     &  $0.532 \pm 0.016$  
  & $0.546 \pm 0.019$    & $0.482 \pm 0.020$ 
  & $0.541 \pm 0.032$    &  $0.441 \pm 0.045$ \\

 $<\! \tau \!>$    & $0.405 \pm 0.013$     &   $0.396 \pm 0.012$    
   & $0.361 \pm 0.016$  & $0.406 \pm 0.014$  
   &  $0.395 \pm 0.016$ &  $0.425 \pm 0.015$  \\

                                                        \\

 $<\! \log a \!>$  & $0.006 \pm 0.003$      &   $0.023 \pm 0.003$      
            & $0.001 \pm 0.005$ & $0.000 \pm 0.004$ 
            & $0.007 \pm 0.008$   & $-0.030 \pm 0.008$  \\

 $<\! {\rm i} \!>$       &  $14\fdg6 \pm 4\fdg0$      &  $25\fdg9 \pm 2\fdg4$  
  & $32\fdg6 \pm 1\fdg6$ & $51\fdg0 \pm 1\fdg0$ 
  & $73\fdg2 \pm 0\fdg7$ &  $91\fdg2 \pm 0\fdg2$     \\

 $<\! \omega \!>$  & $263\fdg2 \pm 56\fdg4$     &  $260\fdg4 \pm 4\fdg9$ 
    & $245\fdg6 \pm 4\fdg5$  & $251\fdg2 \pm 1\fdg8$ 
    & $252\fdg0 \pm 1\fdg6$  & $249\fdg5 \pm 2\fdg6$     \\

 $< \! \Omega \!>$ &  $ 94\fdg6 \pm 25\fdg1$      &  $108\fdg8 \pm 5\fdg4$  
   & $123\fdg9 \pm 4\fdg4$  & $117\fdg7 \pm 1\fdg4$ 
   & $119\fdg5 \pm 0\fdg6$  & $118\fdg1 \pm 0\fdg1$    \\

                                                                         \\

 $<\! \log {\cal M} \!>$  & $0.004 \pm 0.026$   & $0.003 \pm 0.026$  
            & $-0.087 \pm 0.037$  & $0.012 \pm 0.028$ 
            & $-0.021 \pm 0.035$ &  $0.069 \pm 0.029$         \\

                                                                  \\

\cline{1-7}

\hline
\hline

\end{tabular}

\end{minipage}
\end{table*}

\acknowledgement

I thank D.J.Mortlock for a helpful discussion on nuisance parameters
and profile likelihoods. I am obliged to the referee for suggestions that
improved the paper.

\end{document}